\documentclass[lettersize,journal]{IEEEtran}
\usepackage{amsmath,amsfonts}
\usepackage{algorithmic}
\usepackage{algorithm}
\usepackage{array}
\usepackage[caption=false,font=normalsize,labelfont=sf,textfont=sf]{subfig}
\usepackage{textcomp}
\usepackage{stfloats}
\usepackage{url}
\usepackage{verbatim}
\usepackage{graphicx}
\usepackage{cite}
\usepackage{lipsum}
\usepackage{lineno}
\usepackage{siunitx}
\usepackage{todonotes}
\newcommand{\restr}[2]{{
  \left.\kern-\nulldelimiterspace
  #1
  \vphantom{\big|}
  \right|_{#2}
}}
\DeclareSIUnit\electron{e^-}
\begin{document}

\title{Extending the field of view in modulation-based X-ray phase microtomography}

%\thanks{Manuscript received April 19, 2021; revised August 16, 2021.}}
\author{%
Dominik John\textsuperscript{*},
Junan Chen\textsuperscript{*},
Christoph Gaßner,
Sara Savatović,
Lisa Marie Petzold,
Sami Wirtensohn,
Mirko Riedel,
Jörg U. Hammel,
Julian Moosmann,
Felix Beckmann,
Matthias Wieczorek,
and Julia Herzen%
\thanks{*Dominik John and Junan Chen contributed equally to this work.}%
\thanks{Dominik John, Sami Wirtensohn, Jörg U. Hammel, Julian Moosmann, and Felix Beckmann are with the Institute of Materials Physics, Helmholtz-Zentrum hereon, Max-Planck-Straße 1, 21502 Geesthacht, Germany.}%
\thanks{Junan Chen and Matthias Wieczorek are with ImFusion GmbH, Agnes-Pockels-Bogen 1, 80992 München, Germany.}%
\thanks{Dominik John, Junan Chen, Christoph Gaßner, Sara Savatović, Lisa Marie Petzold, Sami Wirtensohn, Mirko Riedel, and Julia Herzen are with the Research Group Biomedical Imaging Physics, Department of Physics, TUM School of Natural Sciences, Technical University of Munich, 85748 Garching, Germany, and also with the Munich Institute of Biomedical Engineering, Technical University of Munich, 85748 Garching, Germany.}%
\thanks{Sara Savatović is also with the Department of Physics and Astronomy "Galileo Galilei", University of Padua, 35131 Padua, Italy.}%
\thanks{Corresponding author: Dominik John (e-mail: dominik.john@tum.de).}%
}

%{Shell \MakeLowercase{\textit{et al.}}: A Sample Article %Using IEEEtran.cls for IEEE Journals}


\maketitle

\begin{abstract}
Recent advances in propagation-based phase-contrast imaging, such as hierarchical imaging, have enabled the visualization of internal structures in large biological specimens and material samples. However, modulation-based techniques, which provide quantitative electron density information, face challenges when imaging larger objects due to stringent beam stability requirements and detector distortions. Extending the field of view of these methods is crucial for obtaining comparable quantitative results across beamlines and adapting to the smaller beam profiles of fourth-generation synchrotron sources.
We introduce a novel image processing technique combining an eigenflat optimization with deformable image registration to address the challenges and enable quantitative high-resolution scans of centimeter-sized objects with multiple-micrometer resolution. We demonstrate the potential of the method by obtaining an electron density map of a rat brain sample 15$\,$mm in diameter despite the limited horizontal field of view of 6$\,$mm of the beamline. This showcases the technique's ability to significantly widen the range of applications of modulation-based techniques in both biological and materials science research.
\end{abstract}

\begin{IEEEkeywords}
Principle Component Analysis, Deformable Image Registration, Eigenflat Optimization
\end{IEEEkeywords}

\section{Introduction}
\IEEEPARstart{T}{he} application of phase-contrast imaging at synchrotron sources has significantly enhanced the exploration of internal structures in materials and biological specimens, offering detailed insights, for example, into the mechanisms of disease and material structure \cite{topperwien2018three, muller2022three, taphorn2022x}. A range of different approaches has been developed and successfully applied to extract the phase information imparted onto the X-ray wavefront by the sample, including e.g. propagation-based imaging\cite{paganin2002simultaneous, reichmann2023human}, ptychographic methods \cite{pfeiffer2018x}, Fokker-Planck based approaches \cite{paganin2019x, leatham2024x}, and speckle-based imaging \cite{zdora2018state} -- the latter being more generally referred to as modulation-based imaging to include structured diffusors \cite{quenot2022x, savatovic2025high}. Methods for imaging large objects at the highest possible resolution are currently an active area of research, because they can provide a better understanding of biological processes and material properties \cite{du2021upscaling, ruben2022full, allan2025offset}. They are becoming even more relevant with the advent of fourth-generation synchrotron sources, which typically feature very small angular source widths \cite{chapman2023fourth}, leading to smaller fields of view at the same measurement distance. Recently, hierarchical propagation-based imaging has successfully been used to obtain morphological information of different human organs down to the micrometer level \cite{walsh2021imaging}. For this purpose, projections are acquired at multiple positions and concatenated before reconstruction to yield information on the complete organ.

A similar approach is currently difficult to replicate for modulation-based techniques, but would allow retrieving quantitative information on the electron density of the large object while also not requiring the single-material assumption (i.e., a fixed $\delta/\mu$-ratio). In these techniques, a modulator is placed in the beam path to create a reference pattern, which the object then distorts. Processing approaches such as cross-correlation \cite{berujon2012x} or least-squares minimization (e.g. Unified Modulated Pattern Analysis (UMPA) \cite{Zdora2017}) then allow retrieving three different signals -- attenuation, phase-contrast and dark field -- from multiple tomograms taken at different marker positions. For the purpose of accurate signal retrieval, the beam profile needs to remain constant for both the reference and the sample scan, i.e. the source must remain stable over the time horizon of the measurement -- a condition assumed in conventional flat-field correction \cite{seibert1998flat}. For most synchrotron radiation sources, the beam profile varies significantly over time (e.g. due to beamline component motion or the electron beam orbit control). These discrepancies between the reference image and the conditions during the measurement may cause the modulator (e.g., a grid pattern \cite{Gustschin2021}) to remain visible in the retrieved projections, compromising the image quality and quantitative accuracy of the imaging method.

To address this problem, approaches like a sample/diffusor drift correction for the case of UMPA \cite{savatovic2023helical, savatovic2024extending} and an eigenflat approach have been developed. In the latter approach, a group of acquired flat-field images is decomposed into eigencomponents. A combination of these eigencomponents is then fitted to each projection separately to provide a synthetic flat-field image \cite{Nieuwenhove}. This method has also been successfully applied to phase-contrast scans using modulation-based imaging, but currently requires the presence of a sample-free region within the scan to accurately capture the displacement of the modulator \cite{riedelphd}. When acquiring multiple tomograms at different positions of the same object, only the tomograms capturing the edges of the object include such a sample-free region. For the inner tomograms, the entire field of view is covered by the object and the current eigenflat method is therefore not applicable.

An additional challenge for stitching multiple such tomograms are possible distortions of the projection images due to the time-varying beam profile and the spatially varying detector distortion. Such distortions may not be accurately described by assuming rigid transformations (translation and rotation), as is typically the case for conventional stitching techniques used for panorama creation \cite{stitching1} or medical diagnosis \cite{ADWAN201632}. For this problem, we therefore propose the use of deformable image registration, a technique applied successfully for example in image-guided surgery \cite{RISHOLM2011197} and radiation therapy \cite{radiationtherapy}.

In response to the mentioned challenges faced by modulation-based techniques, we present a novel approach combining an eigenflat method developed for this scenario with deformable image registration. In this way, we allow for an arbitrary expansion of the total field of view by combining multiple tomograms acquired at different positions to cover the entire sample. This means it becomes possible to obtain electron density maps for objects several centimeters in size with multiple-micrometer resolution without strict requirements on the stability of the beam profile or sample over time. Due to the quantitative nature of the method, the results are more comparable over time and at different beamlines. 

As an example application, the method is used to retrieve an electron density map of the full horizontal extent of a rat brain \SI{15}{\milli\meter} in diameter at the beamline P05, operated by Hereon, at PETRA III (\emph{DESY}, Hamburg), despite the limited \SI{6}{\milli\meter} horizontal FOV. This scan utilizes Talbot array illuminators as modulators; however, we also apply the method to data acquired with a sandpaper modulator, as described in the appendix.

\section{Methodology}
\subsection{Scan acquisition}
\begin{figure}
    \centering
    \includegraphics[width=\linewidth]{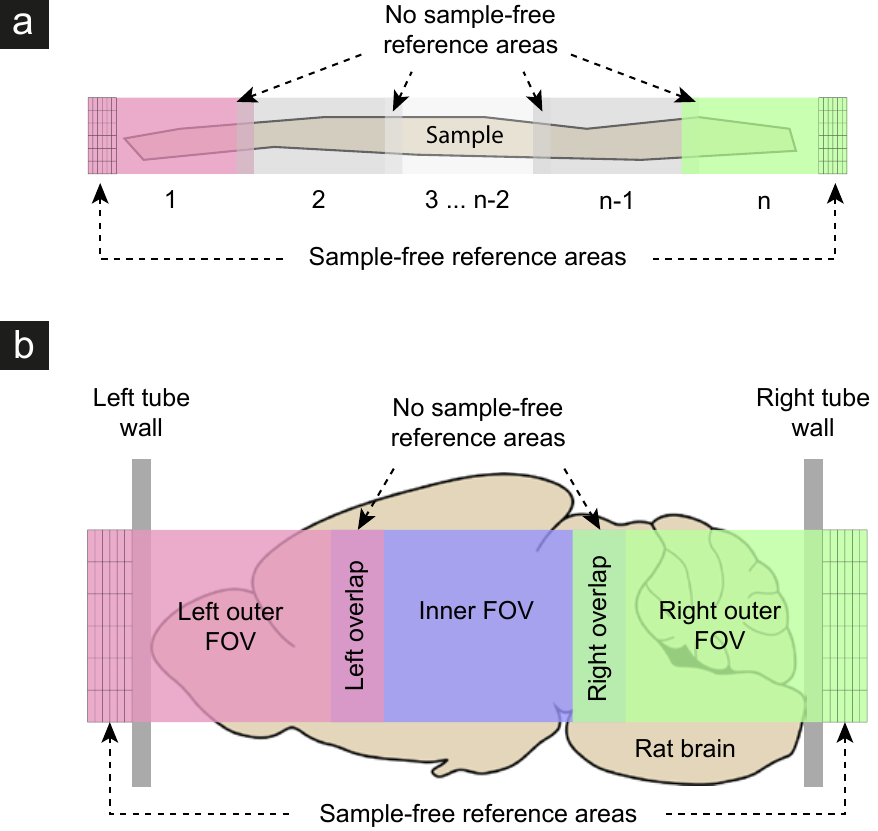}
    \caption{Illustration of the image acquisition strategy. The sample rotates around the same axis for every scan, but different parts of the sample are recorded in succession from left to right using the available FOV. While the conventional eigenflat optimization can be performed for the left-most and right-most scans due to the presence of a sample-free region, this is not the case for the inner scan(s). (a) shows the approach for $n$ overlapping scans, in which only scan 1 and scan $n$ contain a sample-free reference region. (b) illustrates the special case of $n = 3$.}
    \label{fig:scan-sketch}
\end{figure}

\begin{figure*}[!t]
    \centering    
    \includegraphics[width=\textwidth]{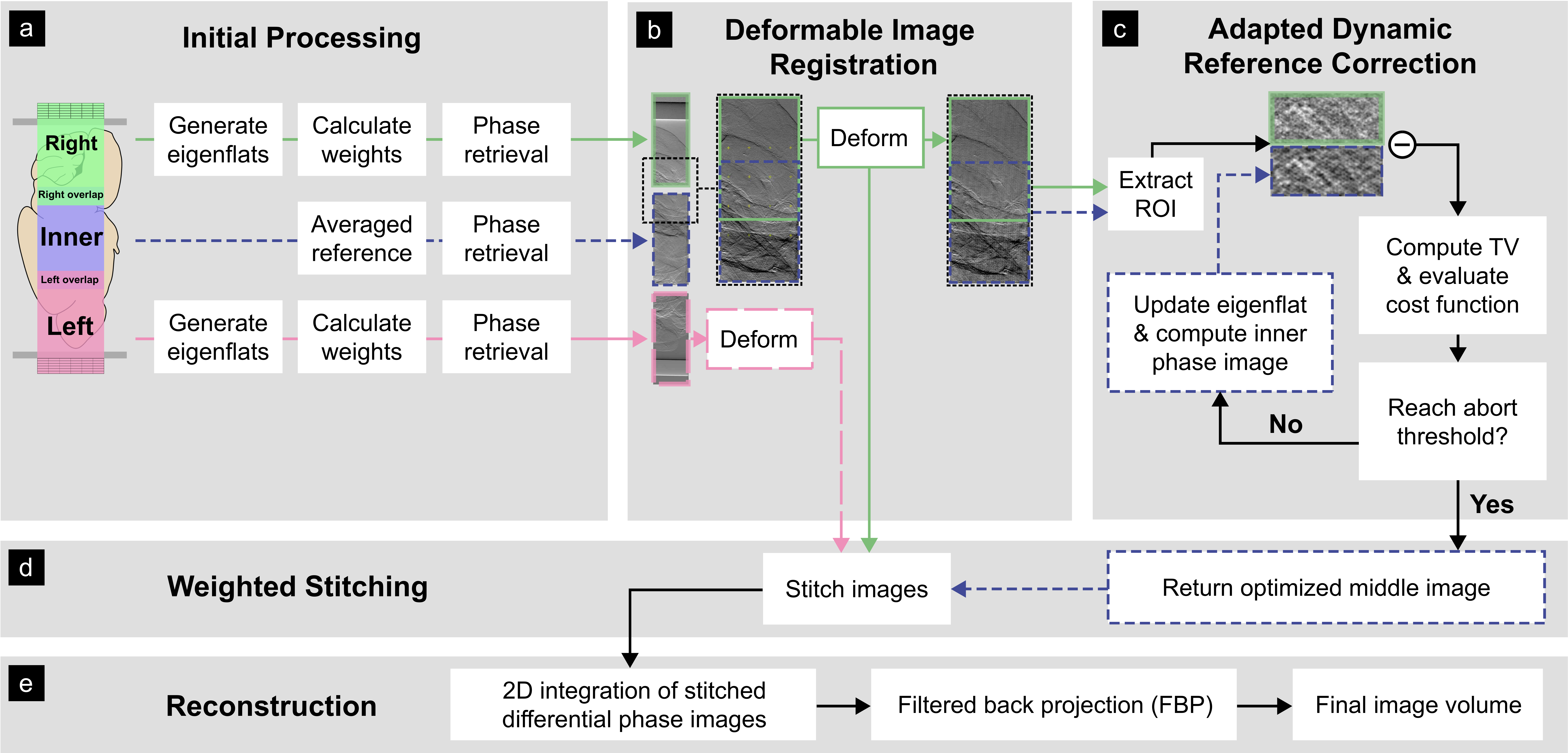}
    \caption{Illustration of the scan processing pipeline. (a) For the left and right scan data, synthetic reference images are created by matching eigenflats to the sample-free region. For the middle scan, the mean reference images are used. In all three cases, initial differential phase images are retrieved.
    (b) The differential phase images are used for deformable image registration, where the right image is matched to the middle image. A B-spline weighted free-form deformation model is defined by the control point grid, which is displayed as the yellow markers in the figure. The same procedure is applied to the left image together with the middle image (not illustrated).
    (c) A ROI is extracted from the overlapping region both from the middle and the right scan. The weights for the synthetic reference of the middle scan are adjusted iteratively until a satisfactory match between the differential phase images of the middle and right scan is achieved.
    (d) The calculated weights of the eigenflats for the middle scan are used to retrieve the final corrected middle phase image. This image is stitched with the previously deformed left and right phase images to yield the final projection image. The procedure is repeated for all projections. (e) Integration is performed on the stitched differential phase images, leading to the final phase projections of the object. After application of a filtered back projection, the three-dimensional electron density volume is obtained.}
    \label{fig:procedure}
\end{figure*}

The general scanning strategy for samples that do not fit into a single FOV is illustrated in Figure \ref{fig:scan-sketch}(a). The approach is applicable to any number of recorded subregions, but is described for the case of $n = 3$ scans as visualized in Figure \ref{fig:scan-sketch}(b) in the following. At first, full tomographic data sets of the left outer, inner, and right outer regions of the sample are acquired separately and in this order, resulting in three sets of computed tomography (CT) projections. The iso-centers of these scans are set fixed and the difference between the scans is equivalent to a change in the horizontal offset of the detector and source. The horizontal offsets of the detector and source are chosen such that there is a small overlap between projection images from different sample parts, indicated as left and right overlap respectively in Figure \ref{fig:scan-sketch}(b). The overlap size should be chosen on the order of $\SI{10}{\percent}$ of the single field of view.

\subsection{Processing}
The initial processing is shown in Figure \ref{fig:procedure}(a). The outer projection images are acquired such that there are sample-free areas at the borders. The eigenflat-field correction, as first described in Ref.~\cite{Nieuwenhove}, is then applied: Given a set of $M$ recorded flat-field images $\boldsymbol{F}=\left(\boldsymbol{f}_1, \ldots, \boldsymbol{f}_M\right)$ with $N$ pixels each, the mean of all flatfields $\overline{\boldsymbol{f}}$ is first subtracted from each to obtain the centered flatfield matrix $\boldsymbol{A} \in \mathbb{R}^{N \times M}$:

\begin{equation}
    \boldsymbol{A}=\left(\boldsymbol{f}_1-\overline{\boldsymbol{f}}, \ldots, \boldsymbol{f}_M-\overline {\boldsymbol{f}}\right).
\end{equation}

The eigenflat-fields ${\boldsymbol{u}_k}$ are defined as the eigenvectors of the covariance matrix $\boldsymbol{C} = 
\boldsymbol{A} \boldsymbol{A}^T$. For each image, a matching flatfield is calculated by least-squares optimization of the difference between pixels in a sample-free region of the image and the corresponding region in the flatfield $\widehat{\boldsymbol{f}}_j$, where $\hat{w}_{j k}$ are the weights subject to optimization: 

\begin{equation}
\label{eq:weights}
\widehat{\boldsymbol{f}}_j=\overline{\boldsymbol{f}}+\sum_{k=1}^K \hat{w}_{j k}\boldsymbol{u}_k.
\end{equation}

The optimized flats ${\boldsymbol{f}_j}$ are used in the modulation-matching algorithm to obtain the differential phase images in $x$- and $y$-direction. In a first step, the inner projection images of the sample are also input into the modulation-matching algorithm together with a mean reference image created by computing the average of all originally recorded flat-field images.

The data are then processed using an implementation of UMPA \cite{de2023high}, yielding differential phase images in two orthogonal directions for all three parts of the image. The inner projections contain grid-shaped artifacts due to the time-varying nature of the beam, which does not allow to remove the modulator pattern completely using a mean flat approach. Before an adapted eigenflat correction can be applied to resolve this problem, the three image parts first must be registered precisely.

\subsection{Registration}
The projection images may exhibit structural changes due to detector distortion at different positions. These effects, together with different beam profiles at various positions within the FOV, mean that the pair of images to be registered may not be perfectly aligned if the images are merely rigidly shifted based on the horizontal offsets. To cope with these potentially complicated deformations, we apply a deformable image registration algorithm (\emph{ImFusion GmbH}, Munich, Germany \cite{imfusion}). A free-form deformation (FFD) model \cite{FFD} with B-Splines interpolation is used. The deformation model is modified by a control point grid. For each pixel inside the grid, its displacement is determined by the displacements of the 4 nearest control points. The resulting deformation field is continuous. Aiming to find the optimal deformation model, the deformable registration is equivalent to an optimization problem. During the optimization, the displacements of the control points are the variables to be optimized, and the similarity metric between the images is to be maximized.

The deformable registration is demonstrated in Figure \ref{fig:procedure}(b). The inner image is assumed to be stationary and left undeformed, while the outer regions receive deformation models. This particular choice creates a computational advantage for the adapted eigenflat-approach introduced in the next subsection. The deformable image registration is then performed on either the differential phase images in the $x$- or the $y$-direction for all projections. We perform the registration consecutively for all the projections. To encourage smoothness between adjacent projections, as well as improve computational efficiency, we use the optimal control point displacements from the projection at the current angle as the initial guess of the projection at the next angle.

\subsection{Dynamic flat-field correction}
In the next step, eigenflats ${\boldsymbol{u}_k}$ for the inner scan are calculated. However, the matching flat-field images cannot be obtained directly via the approach described in the previous section: The optimal weights ${\hat{w}_{j k}}$ (see Eq. \eqref{eq:weights}) cannot be computed directly because no sample-free region for matching is present. A different optimization problem targeting the optimal weights for the eigenflats of the inner projection is therefore set up, as shown in Figure \ref{fig:procedure}(c).

In each optimization step, a flat-field image for the inner projection is calculated based on the current guess of the eigenflat weights ${\hat{w}_{j k}}$ and used to retrieve differential phase images $\frac{\partial \Phi}{\partial x} (x,y)$ and $\frac{\partial \Phi}{\partial y} (x,y)$ using the phase-retrieval algorithm (e.g.~UMPA), where $\Phi$ is the phase of the measured object. Only one of the two signals is required for the optimization to follow, so we define the pixel value as:
\begin{equation}
I(x,y) \equiv \frac{\partial \Phi} {\partial x} (x,y),
\end{equation}
for conciseness. We can then use the pixel values $I_{\text{inner}}(x,y)$ and $I_{\text{outer}}(x,y)$ from the inner and right outer scans respectively to create a difference image $I_{\text{D}}(x,y)$. We note here that $I_{\text{outer}}(x,y)$ refers to the outer image after application of the deformable registration. Inside a region $\Omega$, defined as the overlap region(s) between the inner and outer scans, $I_{\text{D}}(x,y)$ is approximated as follows:
\begin{equation}
I_{\text{D}}(x,y) = I_{\text{inner}}(x,y) - I_{\text{outer}}(x,y) \approx I_{\text{MP}}(x,y) + N(x,y).
\end{equation}
Here, we have decomposed the difference between the two images into a residual modulation pattern $I_{\text{MP}}(x,y)$ and a pixel-dependent noise term $N(x,y)$. This decomposition is valid provided that a sufficiently good match has been achieved between the inner and outer image regions during registration, such that any remaining difference arises from the modulator pattern or noise rather than displaced image features. In an ideal case in which a perfectly matching flat-field has been used for the correction, the term $I_{\text{MP}}(x,y)$ should equal 0, since the modulator pattern is successfully removed by the modulation-tracking algorithm. Neglecting the noise term, we may therefore use the homogeneity of the difference image $I_{\text{D}}(x,y)$ as a measurement of the flat-field correction quality. We choose the total variation (TV) for this purpose, since it is a well-established metric for evaluating the spatial variation and the smoothness of an image \cite{TV}. For a two-dimensional image \(\boldsymbol{v}\), it is defined as
\begin{equation}
\text{TV}(\boldsymbol{v}) = \sum_{(x,y)} |\boldsymbol{v}_{x+1,y} - \boldsymbol{v}_{x,y}| +  |\boldsymbol{v}_{x,y} - \boldsymbol{v}_{x,y+1}|,\
\end{equation}
where \(x\) and \(y\) denote the pixel indices in the horizontal and vertical direction of the image. The TV of the difference image is then computed and returned as the value of the cost function of the optimization. The cost function is defined as
\begin{equation}
L = \text{TV}\left( \restr{I_{\text{inner}}(x,y)}{\,\Omega} - \restr{I_{\text{outer}}(x,y)}{{\,\Omega}}\right),
\end{equation}
which is eventually minimized to find the optimal weights $\hat{w}$, as mentioned in Eq. \eqref{eq:weights}.

Using the determined weights for the eigenflats, new flat-field images are generated and used to retrieve the final, corrected differential phase images of the middle scan. The left outer, inner, and right outer differential phase-contrast images are then transformed and deformed using the information from the horizontal offsets and the deformation model previously determined by the deformable registration. 

\subsection{Stitching}
Finally, we blend the overlapping regions of the images using a distance-weighted sum
\begin{equation}
I_{\text{blended}}(x,y) = \alpha(x)\, I_{\text{inner}}(x,y) + \left(1 - \alpha(x)\right)\, I_{\text{outer}}(x,y),
\end{equation}
where
\begin{equation}
    \alpha(x) = \frac{\Delta r - l(x)}{\Delta r}.
\end{equation}
Here, the intensity $I_{\text{blended}}(x,y)$ of a pixel in the overlapping region is computed from the pixel intensity of the inner image $I_{\text{inner}}(x,y)$ and the right outer image $I_{\text{outer}}(x,y)$. The weight $\alpha(x)$ depends on the width of the overlapping region $\Delta r$ and the distance of the current pixel to the left boundary of the overlapping region $l(x)$.

The three projections are stitched using the same method to form one projection with an enlarged FOV. The stitched differential phase images in the $x$- and $y$-direction are then integrated using a 2D Fourier integration approach \cite{Kottler-07} to retrieve the final integrated phase signal. This signal is then reconstructed into a three-dimensional electron density map using a Feldkamp-Davis-Kress (FDK) algorithm, as implemented in the \emph{Xaid} software (\emph{Mitos GmbH}, Garching, Germany).

\subsection{Cascading approach for arbitrary number of projections}
\label{sec:cascading-section}
We note here that the proposed method is applicable to an arbitrary number of projections instead of three, as shown in Figure \ref{fig:scan-sketch}(b). The correction will then be performed consecutively from outer to inner projections. For a projection $n$, where $n > 2$, the overlapping region with the previously corrected projection $n-1$ is considered. The cost function for the optimization on projection $n$ is defined as
\begin{equation}
\label{eq:cascading}
L_n = \text{TV}\left(\restr{I_{n}(x,y)}{\,\Omega} - \restr{I_{n-1}(x,y)}{\,\Omega}\right),
\end{equation}
where $\Omega$ again refers to the overlapping region between the two scans.

\section{Experiments}
\subsection{Beamline parameters}
The imaging beamline P05 uses X-rays from an undulator located $d = \SI{85}{\metre}$ from the microtomography endstation. A double-crystal silicon monochromator provides a highly monochromatic beam ($\Delta E / E \approx 10^{-4}$) with an energy of \SI{30}{\kilo\eV}. The beam is approximately parallel with a beam divergence of $28 \times \SI{4.0}{\micro\radian\squared}$ at \SI{10}{\kilo\eV} \cite{wilde2016micro}. A CMOS camera based on the CMOSIS CMV 20000 sensor \cite{lytaev2014characterization}, developed in collaboration with the Karlsruhe Institute of Technology (KIT), is positioned \SI{620}{\milli\metre} behind the sample. It features an array of $5120 \times 3840$ pixels with a physical pixel size of \SI{6.4}{\micro\metre}, which leads to an effective pixel size of \SI{1.28}{\micro\metre} using a 5-fold magnification lens. The objective is coupled to a \SI{100}{\micro\metre}-thick $\text{CdWO}_4$ scintillator.

The modulator is a two-dimensional phase-shifting grating referred to as a Talbot array illuminator \cite{Gustschin2021}. It is positioned \SI{175}{\milli\metre} in front of the sample. The modulator has a period of \SI{10}{\micro\metre} and introduces a phase shift of $\Delta \phi = 2\pi/3$ to parts of the beam in a hexagonal lattice structure. The duty cycle, which specifies the ratio of hole spacing to hole radius, is $1/3$. The modulator was fabricated on \SI{200}{\micro\metre}-thick silicon wafers using deep reactive ion etching to produce round holes of suitable depth. At fractional Talbot distances of $\frac{1}{6} d_\text{T}$, the modulators exhibit a focusing effect with a theoretical compression ratio of 1:3 in each direction, allowing high visibility to be achieved. The grating is stepped using a piezoelectric stepper capable of movement in the two directions orthogonal to the beam path.

\subsection{Sample}
The sample investigated during this work is a whole rat brain. Animal housing and organ removal were carried out at Helmholtz-Zentrum Munich in accordance with the European Union guidelines 2010/63. The procedure was conducted in compliance with the ethical standards of the institution and approved by the responsible governmental body. Specifically, the experiments were performed under the license number ROB-55.2-2532.Vet\_02-21-133. After removal, the sample was fixed in formaldehyde. The sample was then dehydrated using an ethanol series starting with \SI{50}{\percent} (vol/vol) ethanol per water up to \SI{100}{\percent} (vol/vol) ethanol in steps of \SI{10}{\percent} for one hour each. Afterwards, the sample was stained with an eosin stain on an ethanol basis, using \SI{0.6}{\percent} (wt/vol) eosin in \SI{100}{\percent} ethanol. The ethanol-based staining was used due to the high lipid ratio in the brain tissue. The brain contains a MENX-associated mutation \cite{pellegata2012menx}.

\subsection{Measurement protocol}
\begin{figure}
    \centering
\includegraphics[width=\linewidth]{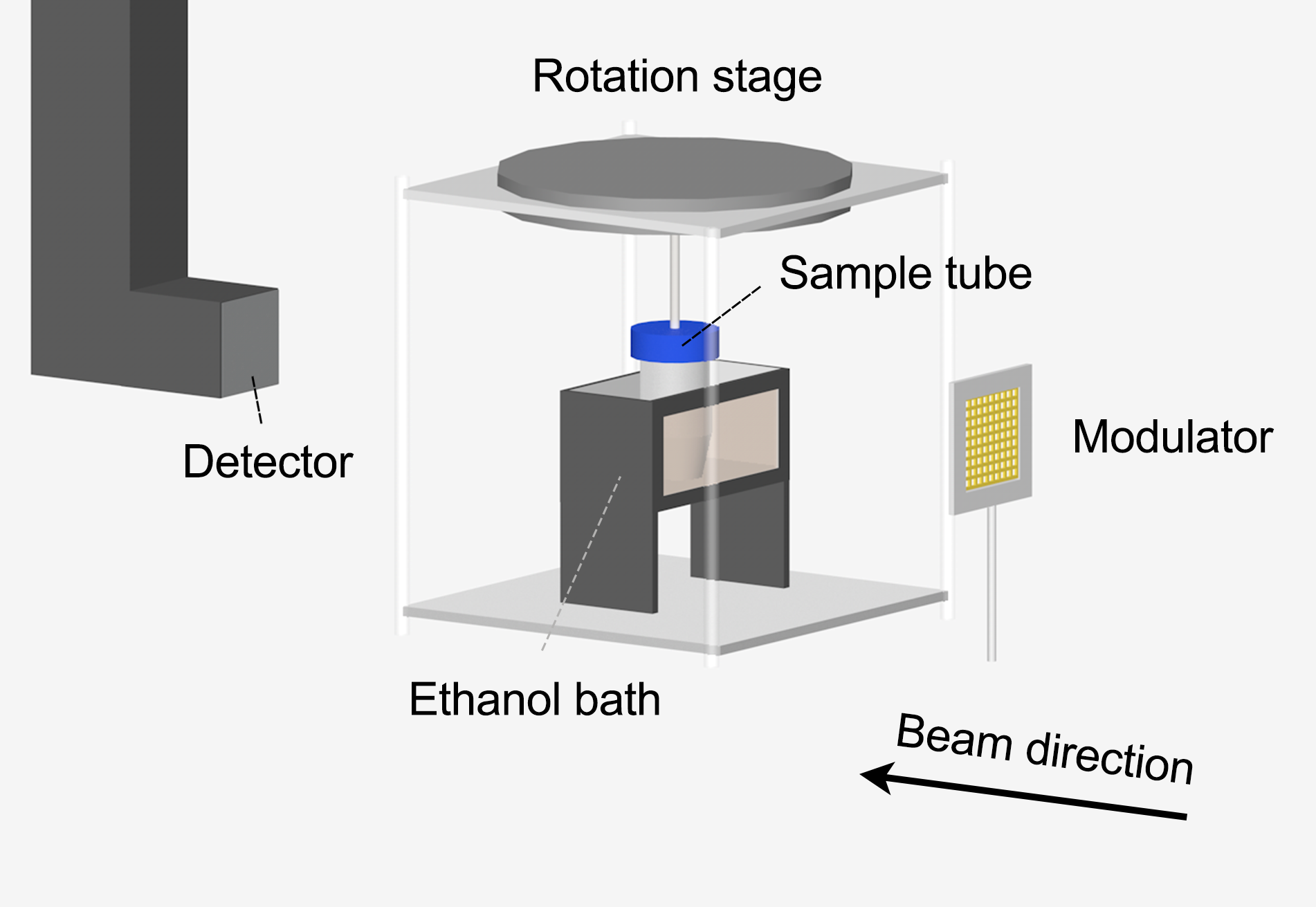}
    \caption{Experimental setup used for the rat brain scan. The beam first passes the modulator, a Talbot array illuminator, before entering the ethanol bath through a Kapton foil window. It then impinges on the sample, which is supported from the top using a rotation stage, and leaves the bath through another Kapton window. Shortly thereafter, it reaches the detector, which consists of a scintillator that is coupled to a CMOS camera using a magnifying lens (not depicted).}
    \label{fig:hanging-axis}
\end{figure}
The rat brain is imaged using a hanging-rotation-axis approach, as illustrated in Figure \ref{fig:hanging-axis}. The sample is placed inside a tube with a diameter of \SI{15}{\milli\metre} and attached to a rotation stage at the top. To minimize variations in electron density, the tube is submerged in a bath filled with ethanol, which prevents the strong changes in phase that would occur if measured against air. This setup also maintains a consistent dynamic range for the phase retrieval method across all scans, which would not be the case if some regions contain air while the inner region is fully occupied by the sample. At the high radiation doses encountered at synchrotron-radiation sources, ethanol is preferred over water because water is prone to radiolysis. This leads to the formation of gas bubbles that can displace tissue during measurements, thereby increasing phase variations.

The scans are recorded at an energy of \SI{30}{\kilo\eV} using a continuous rotation mode; the position of the modulator is moved after each 180$^{\circ}$ rotation for a total of 9 stepping positions per scanned FOV. At the beginning and end of each scan, a set of 70 flat-field images is taken. In total, 4000 projections are acquired for each scan with a \SI{100}{\milli\second}-exposure time. In this work, a complete tomogram is first acquired for the left FOV, as depicted in Figure \ref{fig:scan-sketch}, followed by a complete tomogram for the inner and right FOV. Including motor overhead for the modulator movement, the total scan time is approximately \SI{3.25}{\hour}.

\subsection{Algorithm parameters}
The images are downsampled with a factor of 8 to increase the computational efficiency and mitigate the influence of the modulator background on the registration result. The factor was empirically determined to be sufficient for the modulator pattern to disappear. This is required since the modulation background inside the overlapping region is not necessarily the same for the two scans and may therefore hinder the registration algorithm from matching the actual sample features. Considering a tradeoff between computational efficiency and deformation complexity, 16 control points on a $4 \times 4$ grid are assigned at the overlapping region, resulting in 32 degrees of freedom. To emphasize the local motion, the local normalized cross-correlation \cite{LNCC} is used as the similarity metric for the registration.

Note that 8-fold downsampling is applied only during the registration step. After registration, the optimized deformation is applied to the image at the original resolution before the 8-fold downsampling. The deformed image is then passed to the further steps. The geometrical error induced by beam divergence is neglected, given the approximately parallel beam encountered at most synchrotron radiation sources.

For computational efficiency, only a subregion of the overlapping regions $180 \times 240$ pixels in size is used for the optimization. In our experiments, larger regions did not lead to significant improvements in reference matching, while introducing additional computational complexity. Since the target function wraps the phase images that are dynamically retrieved via the phase retrieval optimization, the target function is non-differentiable. We therefore use the derivative-free optimizer BOBYQA \cite{BOBYQA} from the NLopt library \cite{NLopt}. The optimization variables are the weights of eigenflats, they are therefore unitless. The boundary is empirically set to $[-0.5, 0.5]$, the initial step size is 0.1 and the optimization is terminated when the relative change of the cost value falls below $1 \times 10^{-4}$. We use the first three components (eigenflats) for the flat-field generation, which results in a total of 27 optimization variables ($3$ weights $\times \,9$ steps).

\section{Results and discussion}
We ran the image processing pipeline on the high-performance computing cluster of DESY (Hamburg, Germany), with the approximate runtimes for each processing step in Figure \ref{fig:procedure} stated in the following: The matching reference images were calculated separately for each projection in the left and right FOV (Figure \ref{fig:procedure}(a)), taking \SI{1.5}{\second} per projection. The phase retrieval took a further \SI{2.5}{\second} per projection. Both operations were performed concurrently on 100 compute nodes, each equipped with an AMD EPYC 75F3 CPU. The deformable registration (Figure \ref{fig:procedure}(b)) was performed on a single GPU node equipped with an NVIDIA Quadro RTX 8000 and took \SI{2}{\second} per projection pair. The calculation of matching reference images using the adapted eigenflat approach for the middle projections (Figure \ref{fig:procedure}(c)) was again performed in parallel on 100 CPU nodes and required \SI{2.5}{\second} per projection, followed by UMPA phase retrieval with the same runtime as stated for the left and right FOV.

\begin{figure}[ht!]
\centering\includegraphics[width=\linewidth]{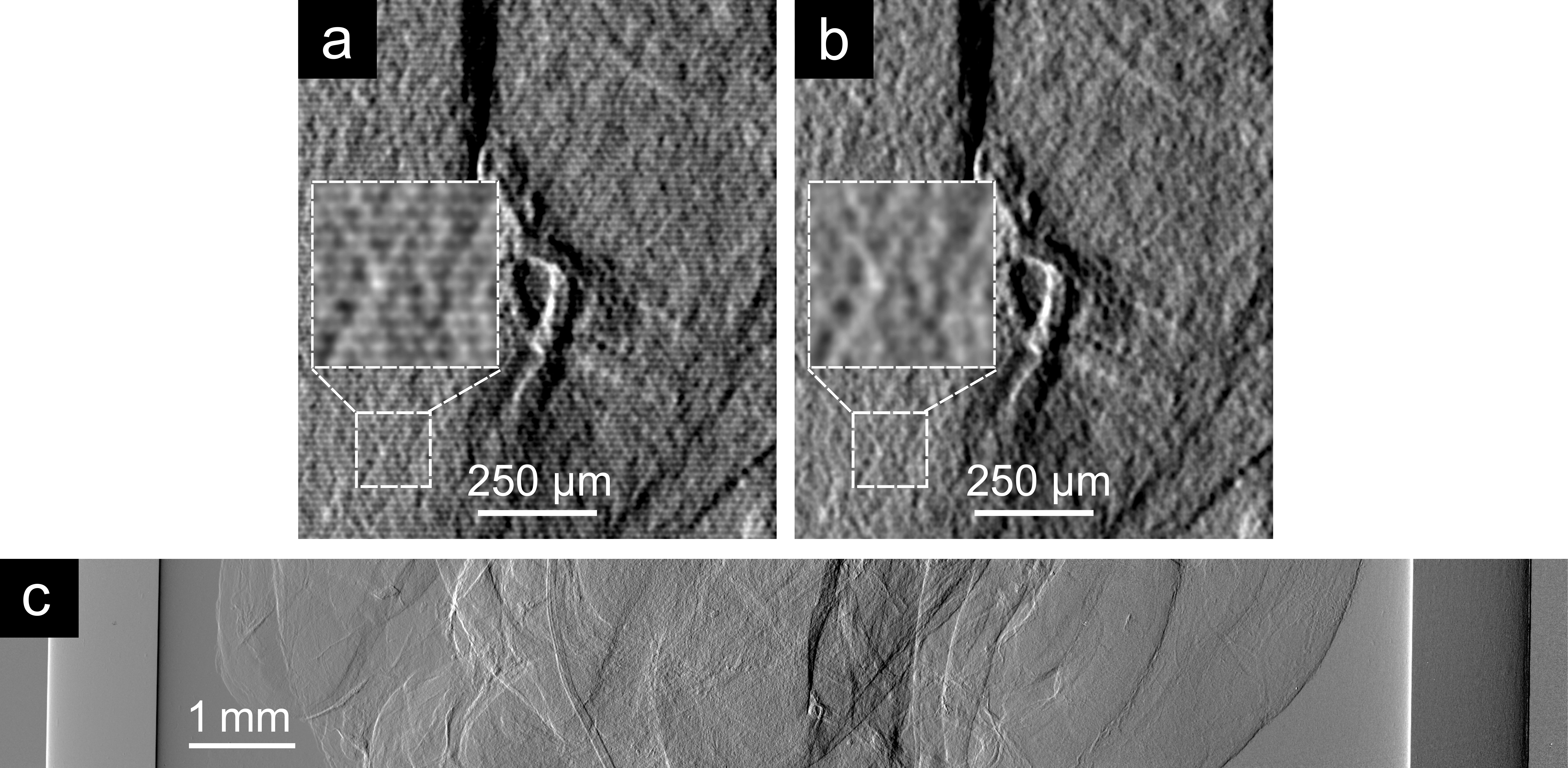}
\caption{Comparison of mean flat and adapted eigenflat approach for a region in the inner scan and the final stitched projection image. (a) Differential phase-contrast projection corrected with mean flat-field images. The tissue is overlayed with artifacts in the form of regular grid patterns, which indicates that the modulation pattern was at different positions during the flat and sample measurement. For this reason, it cannot be removed correctly using the conventional flat-field correction. (b) The same projection corrected with the eigenflats obtained from our proposed method. The appearance of the artifacts is strongly reduced. (c) Example stitched projection image with enlarged FOV showing the differential phase in the $x$-direction.} 
\label{fig:flat-comp}
\end{figure}

Figure \ref{fig:flat-comp} shows a subregion of a projection taken during the inner scan, specifically the differential phase-contrast signal in the $x$-direction. The image is obtained once by a conventional flat-field correction using the mean of the 70 recorded flat-field images (Figure \ref{fig:flat-comp}(a)) and once by our optimization approach (Figure \ref{fig:flat-comp}(b)). While remnants of the 2D grating structure are superimposed onto the anatomical information in (a), this effect is strongly reduced in (b) by the optimized correction. To provide a quantitative comparison of the image quality, we compute the standard deviation and TV within the ROI indicated by the white dashed box. The standard deviation is reduced from 0.067 to 0.062 (\SI{8}{\percent}), while the total variation is reduced from 423.01 to 330.57 (\SI{22}{\percent}). These improvements in both metrics align with the visual evaluation. Note that reducing these metrics to zero is not desired, as this would indicate the removal of actual sample features.

The same correction is repeated for all 4000 projection images of the middle part of the scan. The displacements of the FFD control points as determined by the registration are below \SI{13}{\micro\meter}. This represents the maximum displacement of image pixels in the overlapping region. The three subregions of the sample are then stitched as previously described to obtain a larger combined FOV, as depicted in Figure \ref{fig:flat-comp}(c).

\begin{figure*}[ht!]
\centering\includegraphics[width=1\textwidth]{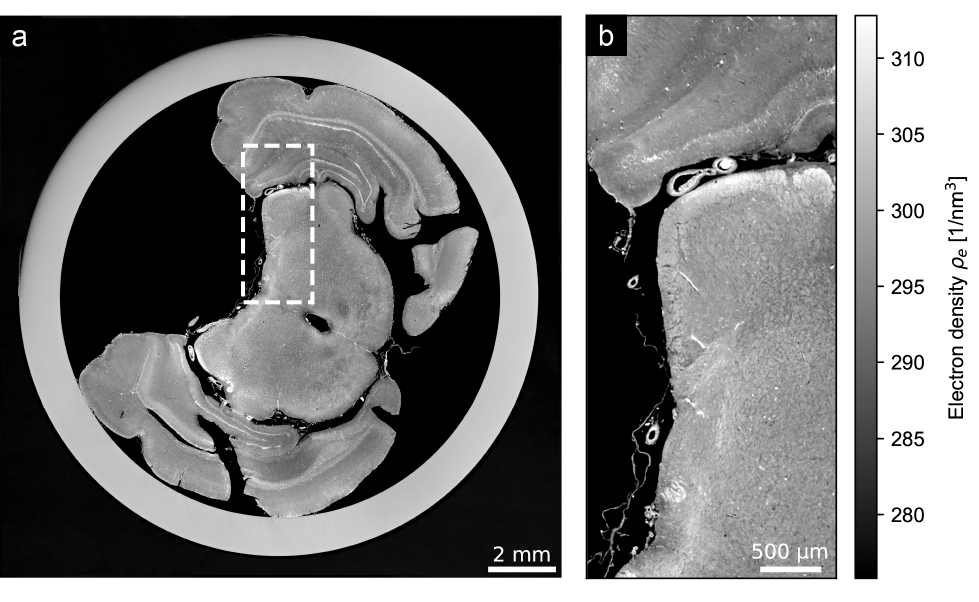}
\caption{(a) Axial slice displaying the entire horizontal extent (\SI{15}{\milli\metre}) of a rat brain inside a tube filled with ethanol. Small differences in electron density between the different tissue types become apparent. Despite being the result of three separate scans with a FOV of \SI{6}{\milli\metre} each, the image gives a uniform impression. (b) Zoom-in region marked by the dotted border in (a).}
\label{fig:ratbrain}
\end{figure*}

After 2D phase retrieval using UMPA with a window size of 3 and reconstruction of the 2D Fourier-integrated phase information using filtered backprojection, an electron density map displaying the entire horizontal extent of the rat brain is obtained. Figure \ref{fig:ratbrain} shows an axial slice of the brain. As becomes apparent, no obvious ring artifacts are visible in the scan, which would otherwise be created by residual grating structures in the projections. Moreover, no obvious artifacts due to the stitching of the three scans are present. At the same time, slight changes in electron density throughout the brain are clearly discernible. The measured electron density is similar to the value of \SI{305}{\per\nano\metre\cubed} previously reported for a human cerebellum \cite{Savatovic:24}.

The overlap in this scan encompassed 1183 pixels in the horizontal direction, but may be reduced by approximately a factor of 6 in size in future scans. This is because only a subregion 180 pixels in horizontal extent of the overlap is used for reference-image matching. Such a reduction in overlap width is desirable to prevent redundancy in the acquired data, leading to unnecessarily prolonged measurement times.

The spatial resolution for the rat brain scan is calculated on a slice of the tomogram using a Fourier ring correlation (FRC) method \cite{saxton1982, vanheel2005} with four image subsets \cite{riedelcomp}. The resolution is determined to be \SI{12.4}{\micro\metre} according to the full-bit and \SI{10.9}{\micro\metre} according to the half-bit criterion. A resolution analysis on different regions of the image confirmed a spatially homogeneous resolution. The final resolution of the scan is affected by multiple parameters: One is the UMPA window parameter, where large windows lead to better noise statistics at the cost of a lower resolution. While the experiment could have been conducted without 2-fold binning of the data, this would have led to a significant increase in computational complexity and scan time due to the additional angles required by the Nyquist sampling theorem.

In the results of the proposed eigenflat optimization shown in Figure \ref{fig:flat-comp}(b), a residual is partially visible as diagonal patterns. This may be attributed to the fact that, due to computational efficiency considerations, we do not utilize all potentially available eigenflats, meaning that not all potentially available eigenflat-field information is considered for the optimization.

\section{Conclusion and Outlook}
In conclusion, we have presented a novel approach that significantly expands the application range of modulation-based imaging techniques. Our method enables the retrieval of small differences in electron density for centimeter-sized objects, given sufficient transmission, by cascading beam reference information inward from outer scans and combining subscans using deformable registration. This advancement is particularly relevant for fourth-generation synchrotron sources, which will feature smaller fields of view, and enables hierarchical scans yielding electron density maps comparable across beamlines, facilitating collaboration and validation of results. The ability to obtain quantitative electron density information for larger samples is a significant step forward in phase-contrast imaging, as it opens up new possibilities for studying complex biological structures and materials at multiple scales. This could be particularly interesting for fields such as biomedical imaging, materials science, and paleontology, where understanding the details of large objects is crucial.

The method was demonstrated for modulation-based imaging but is  applicable in principle to any phase-contrast or attenuation-contrast method requiring accurate flat-field information, provided a suitable optimization metric is available. In this work, we successfully used the total variation (TV) as the optimization metric. In the future, more metrics beyond the native TV should be investigated to simultaneously capture the potential residual low-frequency mismatches in the flat-field images that may occur due to larger-scale shifts of the illumination. While TV is well-suited for characterizing residual modulator patterns in overlap regions, adaptations may be required for region-of-interest scans when samples exhibit pronounced high-frequency spatial components, as the TV metric may inadvertently penalize genuine sample features.

As described in the appendix, our method also allows for measurements of even larger objects than shown here using the cascading approach. However, experimental investigation is needed to determine whether a large number of stitchings $(n \gg 5)$ would lead to an accumulation of error when propagating information from the outermost scans inward. In such a case, however, one may consider correcting the reference information locally for each FOV instead of using the overlap regions, as demonstrated for the ROI scan in the appendix. For $n$ sub-scans, there will be $n-2$ overlap regions that require reference matching, leading to increased computational cost. The precise amount of overlap between scans required likely depends on the experimental conditions and the modulator design, and will therefore need to be determined empirically.

A drawback of our method is the computational intensity of the optimization procedure for each pair of overlapping projections. Future work could explore adapting machine learning-based approaches for synthetic reference generation, such as those developed by Ref.~\cite{grigorev2023flat}, to alleviate this issue for modulation-based phase-contrast imaging as well.

In general, projection stitching approaches require relatively stable samples \cite{bonnin2024multiscale, vo2021data}, as structural changes or movements larger than the pixel size during acquisition can cause irrecoverable artifacts \cite{borisova2021micrometer}. This limitation is particularly problematic for fresh biological tissues that undergo degradation during extended scan times. If sub-volumes can be captured quickly enough compared to the timescale of the sample changes, a volume stitching approach offers the advantage of capturing multiple self-consistent snapshots of the sample \cite{borisova2021micrometer, walchli2021hierarchical}.

Additionally, since a main advantage of modulation-based imaging is that it does not rely on the single-material assumption, future studies should examine samples that strongly violate this assumption to further demonstrate the method's capabilities. Examples include imaging metals embedded in organic tissue \cite{riedelcomp} or other heterogeneous materials. Furthermore, since modulation-based imaging provides a dark-field signal \cite{zdora2018state, savatovic2023multi}, another promising avenue for future research is examining this contrast mechanism for larger samples.

\section*{Acknowledgments}
The authors gratefully acknowledge financial support by the ERC Consolidator Grant (Julia Herzen, TUM, DEPICT, PE3, 101125761) and by the EIC Pathfinder (1MICRON, 101186826). D.J. acknowledges a doctoral scholarship by the Friedrich Naumann Foundation for Freedom funded by the Bundesministerium für Forschung, Technologie und Raumfahrt (BMFTR). We acknowledge DESY (Hamburg, Germany), a member of the Helmholtz Association HGF, for the provision of experimental facilities. Parts of this research were carried out at PETRA III and we would like to thank Fabian Wilde for assistance in using the Beamline P05, operated by the Helmholtz-Zentrum Hereon. This research was supported in part through the Maxwell computational resources operated at DESY. The authors thank Fabio De Marco for fruitful discussions and Natalia S. Pellegata from Helmholtz Munich for providing the rat brain sample.

\section*{Disclosures}
Junan Chen and Matthias Wieczorek are employees of ImFusion GmbH. The remaining authors declare no conflicts of interest.

\section*{Data and code availability} 
Data underlying the results presented in this paper are not publicly available at this time but may be obtained from the authors upon reasonable request. A Python implementation of the algorithm is also available upon request.

{\appendices
\section*{Appendix 1: Cascading reference approach}
We perform a proof-of-concept experiment to verify the feasibility of the cascading approach mentioned in Section \ref{sec:cascading-section}. In this approach, information from the outermost projection is propagated inward to the innermost projection. To test the procedure, we use a measurement that contains three overlapping scans, as shown in Figure \ref{fig:scan-sketch}(b), but do not correct the left outer projection using the conventional approach (i.e., a dynamic flat-field correction using a sample-free reference region). Instead, we assume the left projection to be a scan without a sample-free reference area and use information from the left overlap region, depicted in Figure \ref{fig:scan-sketch}(b), to find the optimal flatfield. After correcting the inner scan as shown in Figure \ref{fig:flat-comp}(b), we find the optimal flatfield for the leftmost scan using the optimization term from Eq.~\eqref{eq:cascading}. Figure \ref{fig:cascade} shows a comparison between the resulting left outer phase image corrected using (a) the mean flatfield and (b) the optimized flatfield obtained via our cascading approach. As becomes apparent, the artifacts caused by leftover modulator structures are reduced in (b). This demonstrates that our approach is also feasible for 3 inner regions, as would be encountered for $n = 5$ regions to be stitched.

\begin{figure}[ht!]
\centering\includegraphics[width=\linewidth]{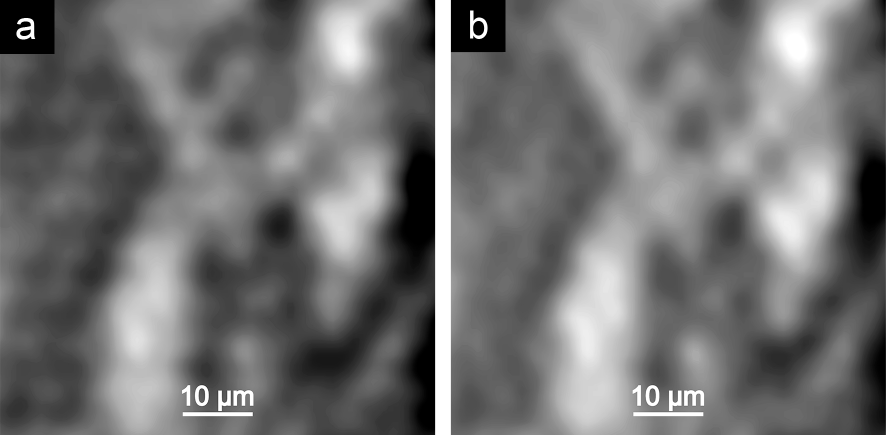}
\caption{Comparison of the mean flat and the cascading eigenflat approach for a region in the
left outer scan. (a) Differential phase-contrast projection corrected with mean flat-field images. (b) The same projection corrected with the eigenflats obtained from the cascading approach. In the optimization, the overlapping region with a previously corrected projection is used as the reference of the current projection.}
\label{fig:cascade}
\end{figure}

\section*{Appendix 2: Sandpaper modulator}
Since the method is applied to Talbot array illuminators in the main section of this work, we want to demonstrate that its working principle is also applicable to sandpaper, a different type of commonly used modulator in modulation-based imaging.

For this purpose, projection data previously acquired in Ref.~\cite{savatovic2023helical} with an effective pixel size of \SI{1.28}{\micro\metre} using a very similar setup is used. The examined object is a phantom sample consisting of polymethylmethacrylate (PMMA), polytetrafluorethylene (PTFE), polychlorotrifluoroethylene (PCTFE), and polyamide nylon-(6,6). It also includes a toothpick and a pipette tip containing polystyrene microspheres.

\label{sec:sandpaper}
\begin{figure}[ht!]
\centering\includegraphics[width=\linewidth]{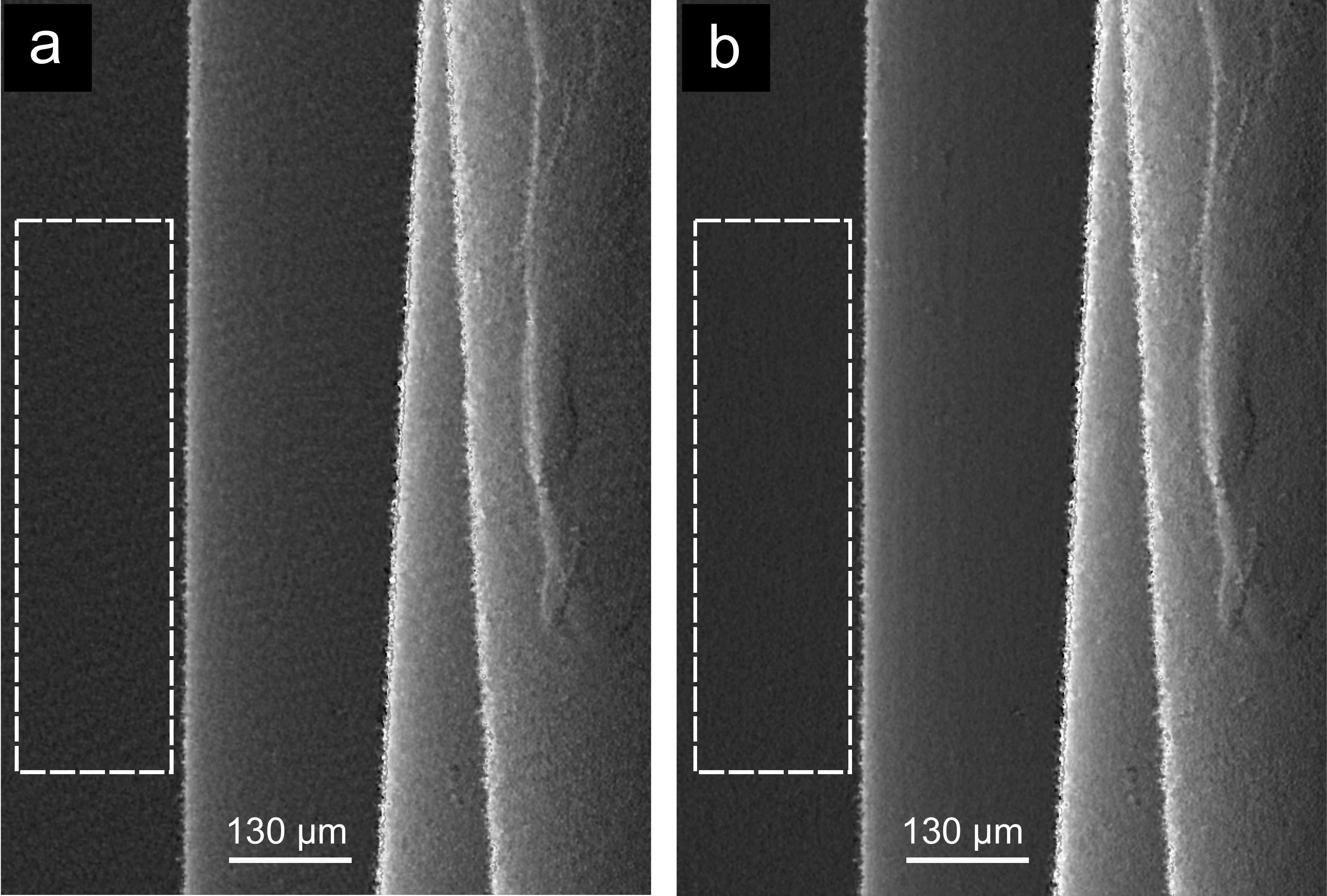}
\caption{Comparison of mean flat and adapted eigenflat approach for an ROI in a scan with sandpaper modulator. (a) Differential phase-contrast projection corrected with mean flat-field images. (b) The same projection corrected with the eigenflats obtained from our proposed method. The standard deviation is computed from the ROI marked by the white dashed box. The appearance of the artifacts is strongly reduced.
} 
\label{fig:sandpaper_result}
\end{figure}
The modulator used for this scan is P1000 sandpaper with a mean particle size of \SI{5.8}{\micro\metre}. The same approach as mentioned in Section \ref{sec:roi-scans} is applied to the acquired data. Figure \ref{fig:sandpaper_result} shows a ROI of the differential phase image retrieved with a mean flatfield (a) and with an optimized flatfield from our method (b). In Figure \ref{fig:sandpaper_result}(b), a reduction in background noise is observed, which marks the removal of the random speckle pattern generated by the sandpaper modulator. Moreover, the standard deviation of the differential phase image decreases from 0.051 to 0.034 (\SI{33}{\percent}). The result visually and quantitatively confirms the feasibility of the proposed method on data acquired with a sandpaper modulator.

\section*{Appendix 3: Region-of-interest scans}
\label{sec:roi-scans}
\begin{figure}[ht!]
    \centering
    \includegraphics[width=\linewidth]{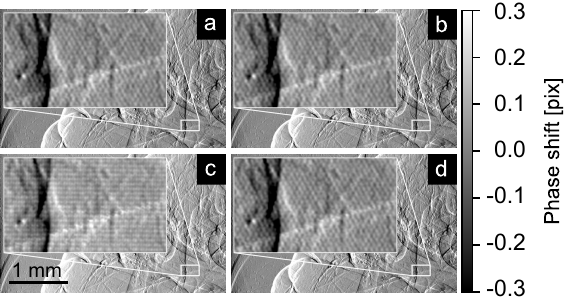}
    \caption{Comparison of the impact on image quality of different approaches to flatfield generation. (a) Projection corrected with a mean flatfield and without eigenflats. The grid pattern remains clearly visible. (b) Projection corrected with eigenflats (5 components) matched using a sample-free region. This approach would not be feasible for a region-of-interest scan, where no such region exists, but rather serves as a comparison. (c) Same approach as (b) but with a matching region inside the sample. This approach leads to strong artifacts. (d) Method proposed in the main section of the paper with a matching region inside the sample and 3 eigenflat components.}
    \label{fig:comparison-roi}
\end{figure}
The reference region used in the proposed optimization is not always available in certain cases, an example being ROI scans. However, a similar strategy is still feasible for the purpose of dynamic flat-field correction. One can adapt the proposed method by evaluating a sample region only. The optimization problem is then turned into minimizing the cost function
\begin{equation}
L = \text{TV}(\restr{\frac{\mathrm{d} \Phi}{\mathrm{d} x}}{\,\Omega}),
\label{eq:roi}
\end{equation}
where $\Omega$ denotes a small region of interest inside the sample from the differential phase image. A ROI with a relatively low variance is selected in order to mitigate the influence of the sample structures on the TV evaluation.
Figure \ref{fig:comparison-roi} provides a comparison of the result from the proposed eigenflat-optimization and the other approaches. Note that the scan itself is not originally a ROI scan since the edge of the sample has been imaged as well, but the behavior of the eigenflat method is simulated by considering a ROI inside the sample region. At the same time, a comparison with approaches using the sample-free reference area becomes possible.

Figure \ref{fig:comparison-roi}(a) is the phase image acquired with a mean flatfield. Panel (b) shows the result using the method proposed in Ref.~\cite{riedelphd}. Note that this method is not applicable for ROI scans given that a sample-free reference region is required, so the result is only provided for comparison. Panel (c) shows the result from the same method as in Figure \ref{fig:comparison-roi}(b), using a sample region instead. Figure \ref{fig:comparison-roi}(d) is the phase image retrieved using the flatfield generated from the proposed optimization in Eq.~\eqref{eq:roi}. In analogy to Figure \ref{fig:comparison-roi}(c), a sample region is used for the optimization. Our result in Figure \ref{fig:comparison-roi}(d) shows a comparable quality to the result in Figure \ref{fig:comparison-roi}(b), however, our method does not require a sample-free reference region as in the other method. Moreover, fewer eigenflat components are needed for our method to generate an image with similar quality (3 vs.~5).}

\bibliographystyle{IEEEtran}
\bibliography{sample}

@article{savatovic2023helical,
  title={Helical sample-stepping for faster speckle-based multi-modal tomography with the Unified Modulated Pattern Analysis ({UMPA}) model},
  author={Savatovi{\'c}, Sara and De Marco, Fabio and Riedel, Mirko and Di Trapani, Vittorio and Margini, Marco and Lautizi, Ginevra and Hammel, J{\"o}rg U and Herzen, Julia and Thibault, Pierre},
  journal={Journal of Instrumentation},
  volume={18},
  number={11},
  pages={C11020},
  year={2023},
  publisher={IOP Publishing}
}

@article{ruben2022full,
  title={Full field X-ray scatter tomography},
  author={Ruben, Gary and Pinar, Isaac and Brown, Jeremy MC and Schaff, Florian and Pollock, James A and Crossley, Kelly J and Maksimenko, Anton and Hall, Chris and Hausermann, Daniel and Uesugi, Kentaro and others},
  journal={IEEE Transactions on Medical Imaging},
  volume={41},
  number={8},
  pages={2170--2179},
  year={2022},
  publisher={IEEE}
}

@article{allan2025offset,
  title={Offset geometry for extended field-of-view in multi-contrast and multi-scale X-ray microtomography of lung cancer lobectomy specimens},
  author={Allan, Harry and Doherty, Adam and Navarrete-Le{\'o}n, Carlos and Jia, Yunpeng and Percival, Charlotte and Hagel, Zoe and Otter, Kate EJ and Khaw, Chuen Ryan and Gowers, Kate and Hall, Helen and others},
  journal={arXiv preprint arXiv:2502.10322},
  year={2025}
}

@inproceedings{savatovic2024extending,
  title={Extending the field-of-view of speckle-based microtomography with the {UMPA} model},
  author={Savatovi{\'c}, Sara and De Marco, Fabio and Riedel, Mirko and Laundon, Davis J and Regner, Mich{\`e}le-Louise and Hammel, J{\"o}rg U and Lewis, Rohan M and Herzen, Julia and Thibault, Pierre},
  booktitle={Developments in X-ray Tomography XV},
  volume={13152},
  pages={131520X},
  year={2024},
  organization={SPIE}
}

@article{de2023high,
  title={High-speed processing of X-ray wavefront marking data with the Unified Modulated Pattern Analysis ({UMPA}) model},
  author={De Marco, Fabio and Savatovi{\'c}, Sara and Smith, Ronan and Di Trapani, Vittorio and Margini, Marco and Lautizi, Ginevra and Thibault, Pierre},
  journal={Optics Express},
  volume={31},
  number={1},
  pages={635--650},
  year={2023},
  publisher={Optica Publishing Group}
}

@article{pellegata2012menx,
  title={{MenX} and {MEN4}},
  author={Pellegata, Natalia S},
  journal={Clinics},
  volume={67},
  pages={13--18},
  year={2012},
  publisher={Elsevier}
}

@article{walsh2021imaging,
  title={Imaging intact human organs with local resolution of cellular structures using hierarchical phase-contrast tomography},
  author={Walsh, CL and Tafforeau, P and Wagner, WL and Jafree, DJ and Bellier, A and Werlein, C and K{\"u}hnel, MP and Boller, E and Walker-Samuel, S and Robertus, JL and others},
  journal={Nature Methods},
  volume={18},
  number={12},
  pages={1532--1541},
  year={2021},
  publisher={Nature Publishing Group US New York}
}

@article{Savatovic:24,
author = {Sara Savatovi\'{c} and Marie-Christine Zdora and Fabio De Marco and Christos Bikis and Margie Olbinado and Alexander Rack and Bert M\"{u}ller and Pierre Thibault and Irene Zanette},
journal = {Biomedical Optics Express},
number = {1},
pages = {142--161},
publisher = {Optica Publishing Group},
title = {Multi-resolution {X}-ray phase-contrast and dark-field tomography of human cerebellum with near-field speckles},
volume = {15},
month = {Jan},
year = {2024},
doi = {10.1364/BOE.502664},
}

@article{vanheel2005,
title = {Fourier shell correlation threshold criteria},
journal = {Journal of Structural Biology},
volume = {151},
number = {3},
pages = {250-262},
year = {2005},
issn = {1047-8477},
doi = {https://doi.org/10.1016/j.jsb.2005.05.009},
author = {Marin {van Heel} and Michael Schatz}
}

@article{saxton1982,
author = {Saxton, W. O. and Baumeister, W.},
title = {The correlation averaging of a regularly arranged bacterial cell envelope protein},
journal = {Journal of Microscopy},
volume = {127},
number = {2},
pages = {127-138},
doi = {https://doi.org/10.1111/j.1365-2818.1982.tb00405.x},
eprint = {https://onlinelibrary.wiley.com/doi/pdf/10.1111/j.1365-2818.1982.tb00405.x},
year = {1982}
}

@article{grigorev2023flat,
  title={Flat-Field Correction of X-Ray Tomographic Images Using Deep Convolutional Neural Networks},
  author={Grigorev, A Yu and Buzmakov, AV},
  journal={Bulletin of the Russian Academy of Sciences: Physics},
  volume={87},
  number={5},
  pages={604--610},
  year={2023},
  publisher={Springer}
}

@misc{chapman2023fourth,
  title={Fourth-generation light sources},
  author={Chapman, Henry N},
  journal={IUCrJ},
  volume={10},
  number={3},
  pages={246--247},
  year={2023},
  publisher={International Union of Crystallography}
}

@article{du2021upscaling,
  title={Upscaling {X}-ray nanoimaging to macroscopic specimens},
  author={Du, Ming and Di, Zichao and G{\"u}rsoy, Dogˇa and Xian, R Patrick and Kozorovitskiy, Yevgenia and Jacobsen, Chris},
  journal={Journal of Applied Crystallography},
  volume={54},
  number={2},
  pages={386--401},
  year={2021},
  publisher={International Union of Crystallography}
}

@article{leatham2024x,
  title={X-ray phase and dark-field computed tomography without optical elements},
  author={Leatham, Thomas A and Paganin, David M and Morgan, Kaye S},
  journal={Optics Express},
  volume={32},
  number={3},
  pages={4588--4602},
  year={2024},
  publisher={Optica Publishing Group}
}

@article{paganin2019x,
  title={X-ray Fokker--Planck equation for paraxial imaging},
  author={Paganin, David M and Morgan, Kaye S},
  journal={Scientific Reports},
  volume={9},
  number={1},
  pages={17537},
  year={2019},
  publisher={Nature Publishing Group UK London}
}

@inproceedings{seibert1998flat,
  title={Flat-field correction technique for digital detectors},
  author={Seibert, James Anthony and Boone, John M and Lindfors, Karen K},
  booktitle={Medical Imaging 1998: Physics of Medical Imaging},
  volume={3336},
  pages={348--354},
  year={1998},
  organization={SPIE}
}

@article{zdora2018state,
  title={State of the art of X-ray speckle-based phase-contrast and dark-field imaging},
  author={Zdora, Marie-Christine},
  journal={Journal of Imaging},
  volume={4},
  number={5},
  pages={60},
  year={2018},
  publisher={MDPI}
}

@article{pfeiffer2018x,
  title={X-ray ptychography},
  author={Pfeiffer, Franz},
  journal={Nature Photonics},
  volume={12},
  number={1},
  pages={9--17},
  year={2018},
  publisher={Nature Publishing Group UK London}
}

@InProceedings{stitching1,
author="Herrmann, Charles
and Wang, Chen
and Bowen, Richard Strong
and Keyder, Emil
and Krainin, Michael
and Liu, Ce
and Zabih, Ramin",
editor="Ferrari, Vittorio
and Hebert, Martial
and Sminchisescu, Cristian
and Weiss, Yair",
title="Robust Image Stitching with Multiple Registrations",
booktitle="Computer Vision -- ECCV 2018",
year="2018",
publisher="Springer International Publishing",
address="Cham",
pages="53--69",
isbn="978-3-030-01216-8"
}

@article{ADWAN201632,
title = {A new approach for image stitching technique using {Dynamic Time Warping} ({DTW}) algorithm towards scoliosis {X}-ray diagnosis},
journal = {Measurement},
volume = {84},
pages = {32-46},
year = {2016},
issn = {0263-2241},
author = {Somaya Adwan and Iqbal Alsaleh and Rasha Majed}
}

@article{RISHOLM2011197,
title = {Multimodal Image Registration for Preoperative Planning and Image-Guided Neurosurgical Procedures},
journal = {Neurosurgery Clinics of North America},
volume = {22},
number = {2},
pages = {197-206},
year = {2011},
note = {Functional Imaging},
issn = {1042-3680},
author = {Petter Risholm and Alexandra J. Golby and William Wells}
}

@article{radiationtherapy,
author = {Brock, Kristy K. and Mutic, Sasa and McNutt, Todd R. and Li, Hua and Kessler, Marc L.},
title = {Use of image registration and fusion algorithms and techniques in radiotherapy: Report of the AAPM Radiation Therapy Committee Task Group No. 132},
journal = {Medical Physics},
volume = {44},
number = {7},
pages = {e43-e76},
year = {2017}
}

@article{reichmann2023human,
  title={Human lung virtual histology by multi-scale x-ray phase-contrast computed tomography},
  author={Reichmann, Jakob and Verleden, Stijn E and K{\"u}hnel, Mark and Kamp, Jan C and Werlein, Christopher and Neubert, Lavinia and M{\"u}ller, Jan-Hendrik and Bui, Thanh Quynh and Ackermann, Maximilian and Jonigk, Danny and others},
  journal={Physics in Medicine \& Biology},
  volume={68},
  number={11},
  pages={115014},
  year={2023},
  publisher={IOP Publishing}
}

@article{paganin2002simultaneous,
  title={Simultaneous phase and amplitude extraction from a single defocused image of a homogeneous object},
  author={Paganin, David and Mayo, Sheridan C and Gureyev, Tim E and Miller, Peter R and Wilkins, Steve W},
  journal={Journal of Microscopy},
  volume={206},
  number={1},
  pages={33--40},
  year={2002},
  publisher={Wiley Online Library}
}

@inproceedings{wilde2016micro,
  title={Micro-{CT} at the imaging beamline {P05} at {PETRA} {III}},
  author={Wilde, Fabian and Ogurreck, Malte and Greving, Imke and Hammel, J{\"o}rg U and Beckmann, Felix and Hipp, Alexander and Lottermoser, Lars and Khokhriakov, Igor and Lytaev, Pavel and Dose, Thomas and others},
  booktitle={AIP Conference Proceedings},
  volume={1741},
  year={2016},
  organization={AIP Publishing}
}

@article{Kottler-07,
author = {C. Kottler and C. David and F. Pfeiffer and O. Bunk},
journal = {Optics Express},
number = {3},
pages = {1175--1181},
publisher = {Optica Publishing Group},
title = {A two-directional approach for grating based differential phase contrast imaging using hard x-rays},
volume = {15},
month = {Feb},
year = {2007},
doi = {10.1364/OE.15.001175},
}

@article{Gustschin2021,
   author = {Alex Gustschin and Mirko Riedel and Kirsten Taphorn and Christian Petrich and Wolfgang Gottwald and Wolfgang Noichl and Madleen Busse and Sheila E. Francis and Felix Beckmann and Jörg U. Hammel and Julian Moosmann and Pierre Thibault and Julia Herzen},
   doi = {10.1364/optica.441004},
   issn = {23342536},
   issue = {12},
   journal = {Optica},
   month = {12},
   pages = {1588},
   publisher = {Optica Publishing Group},
   title = {High-resolution and sensitivity bi-directional x-ray phase contrast imaging using {2D} {Talbot} array illuminators},
   volume = {8},
   year = {2021},
}

@inproceedings{muller2022three,
  title={Three-dimensional imaging and analysis of annual layers in tree trunk and tooth cementum},
  author={M{\"u}ller, Bert and Stiefel, Muriel and Rodgers, Griffin and Humbel, Mattia and Osterwalder, Melissa and von Jackowski, Jeannette A and Hotz, Gerhard and Guadarrama, Adriana A Velasco and Bunn, Henry T and Scheel, Mario and others},
  booktitle={Bioinspiration, Biomimetics, and Bioreplication XII},
  volume={12041},
  pages={98--113},
  year={2022},
  organization={SPIE}
}

@article{taphorn2022x,
  title={{X}-ray Stain Localization with Near-Field Ptychographic Computed Tomography},
  author={Taphorn, Kirsten and Busse, Madleen and Brantl, Johannes and G{\"u}nther, Benedikt and Diaz, Ana and Holler, Mirko and Dierolf, Martin and Mayr, Doris and Pfeiffer, Franz and Herzen, Julia},
  journal={Advanced Science},
  volume={9},
  number={24},
  pages={2201723},
  year={2022},
  publisher={Wiley Online Library}
}

@article{topperwien2018three,
  title={Three-dimensional virtual histology of human cerebellum by X-ray phase-contrast tomography},
  author={T{\"o}pperwien, Mareike and van der Meer, Franziska and Stadelmann, Christine and Salditt, Tim},
  journal={Proceedings of the National Academy of Sciences},
  volume={115},
  number={27},
  pages={6940--6945},
  year={2018},
  publisher={National Academy of Sciences}
}

@phdthesis{riedelphd,
	author = {Riedel, Mirko Philipp},
	title = {Setup Development for High-Resolution Quantitative Phase-Contrast Imaging at a Synchrotron Radiation Source},
	year = {2023},
	school = {Technische Universität München},
	pages = {129},
	language = {en},
	note = {},
	url = {https://mediatum.ub.tum.de/1715444},
}

@article{riedelcomp,
   author = {Mirko Riedel and Kirsten Taphorn and Alex Gustschin and Madleen Busse and Joerg U. Hammel and Julian Moosmann and Felix Beckmann and Florian Fischer and Pierre Thibault and Julia Herzen},
   doi = {10.1038/s41598-023-33788-7},
   issn = {20452322},
   issue = {1},
   journal = {Scientific Reports},
   month = {12},
   pmid = {37117518},
   publisher = {Nature Research},
   title = {Comparing x-ray phase-contrast imaging using a {Talbot} array illuminator to propagation-based imaging for non-homogeneous biomedical samples},
   volume = {13},
   year = {2023},
}

@article{bonnin2024multiscale,
  title={Multiscale Synchrotron Propagation-Based Phase-Contrast X-Ray Tomographic Microscopy at the TOMCAT Beamline: Latest Achievements and Future Plans},
  author={Bonnin, Anne and Lovric, Goran and Marone, Federica and Olbinado, Margie and Schlep{\"u}tz, Christian M and Stampanoni, Marco},
  journal={Synchrotron Radiation News},
  volume={37},
  number={5},
  pages={4--9},
  year={2024},
  publisher={Taylor \& Francis}
}

@article{quenot2022x,
  title={X-ray phase contrast imaging from synchrotron to conventional sources: A review of the existing techniques for biological applications},
  author={Quenot, Laurene and Bohic, Sylvain and Brun, Emmanuel},
  journal={Applied Sciences},
  volume={12},
  number={19},
  pages={9539},
  year={2022},
  publisher={MDPI}
}

@article{savatovic2025high,
  title={High-resolution X-ray phase-contrast tomography of human placenta with different wavefront markers},
  author={Savatovi{\'c}, Sara and Laundon, Davis and De Marco, Fabio and Riedel, Mirko and Hammel, J{\"o}rg U and Busse, Madleen and Salom{\'e}, Murielle and Pascolo, Lorella and Zanette, Irene and Lewis, Rohan M and others},
  journal={Scientific Reports},
  volume={15},
  number={1},
  pages={2131},
  year={2025},
  publisher={Nature Publishing Group UK London}
}

@article{savatovic2023multi,
  title={Multi-resolution {X}-ray phase-contrast and dark-field tomography of human cerebellum with near-field speckles},
  author={Savatovi{\'c}, Sara and Zdora, Marie-Christine and De Marco, Fabio and Bikis, Christos and Olbinado, Margie and Rack, Alexander and M{\"u}ller, Bert and Thibault, Pierre and Zanette, Irene},
  journal={Biomedical Optics Express},
  volume={15},
  number={1},
  pages={142--161},
  year={2023},
  publisher={Optica Publishing Group}
}

@article{vo2021data,
  title={Data processing methods and data acquisition for samples larger than the field of view in parallel-beam tomography},
  author={Vo, Nghia T and Atwood, Robert C and Drakopoulos, Michael and Connolley, Thomas},
  journal={Optics Express},
  volume={29},
  number={12},
  pages={17849--17874},
  year={2021},
  publisher={Optical Society of America}
}

@article{borisova2021micrometer,
  title={Micrometer-resolution {X}-ray tomographic full-volume reconstruction of an intact post-mortem juvenile rat lung},
  author={Borisova, Elena and Lovric, Goran and Miettinen, Arttu and Fardin, Luca and Bayat, Sam and Larsson, Anders and Stampanoni, Marco and Schittny, Johannes C and Schlep{\"u}tz, Christian M},
  journal={Histochemistry and Cell Biology},
  volume={155},
  number={2},
  pages={215--226},
  year={2021},
  publisher={Springer}
}

@article{walchli2021hierarchical,
  title={Hierarchical imaging and computational analysis of three-dimensional vascular network architecture in the entire postnatal and adult mouse brain},
  author={W{\"a}lchli, Thomas and Bisschop, Jeroen and Miettinen, Arttu and Ulmann-Schuler, Alexandra and Hinterm{\"u}ller, Christoph and Meyer, Eric P and Krucker, Thomas and W{\"a}lchli, Regula and Monnier, Philippe P and Carmeliet, Peter and others},
  journal={Nature Protocols},
  volume={16},
  number={10},
  pages={4564--4610},
  year={2021},
  publisher={Nature Publishing Group UK London}
}

@article{berujon2012x,
  title={X-ray multimodal imaging using a random-phase object},
  author={Berujon, Sebastien and Wang, Hongchang and Sawhney, Kawal},
  journal={Physical Review A -- Atomic, Molecular, and Optical Physics},
  volume={86},
  number={6},
  pages={063813},
  year={2012},
  publisher={APS}
}

@inproceedings{lytaev2014characterization,
  title={Characterization of the {CCD} and {CMOS} cameras for grating-based phase-contrast tomography},
  author={Lytaev, Pavel and Hipp, Alexander and Lottermoser, Lars and Herzen, Julia and Greving, Imke and Khokhriakov, Igor and Meyer-Loges, Stephan and Plewka, J{\"o}rn and Burmester, J{\"o}rg and Caselle, Michele and others},
  booktitle={Developments in X-ray Tomography IX},
  volume={9212},
  pages={318--327},
  year={2014},
  organization={SPIE}
}

@article{Zdora2017,
   doi = {10.1103/PhysRevLett.118.203903},
   issn = {10797114},
   issue = {20},
   journal = {Physical Review Letters},
   month = {5},
   pmid = {28581800},
   publisher = {American Physical Society},
   title = {{X}-ray Phase-Contrast Imaging and Metrology through {Unified} {Modulated} {Pattern} {Analysis}},
   volume = {118},
   year = {2017},
}

@article{Nieuwenhove,
   author = {Vincent Van Nieuwenhove and Jan De Beenhouwer and Francesco De Carlo and Lucia Mancini and Federica Marone and Jan Sijbers},
   doi = {10.1364/oe.23.027975},
   issn = {1094-4087},
   issue = {21},
   journal = {Optics Express},
   month = {10},
   pages = {27975},
   pmid = {26480456},
   publisher = {The Optical Society},
   title = {Dynamic intensity normalization using eigen flat fields in {X}-ray imaging},
   volume = {23},
   year = {2015},
}

@Inbook{TV,
author="Caselles, V.
and Chambolle, A.
and Novaga, M.",
title="Total Variation in Imaging",
bookTitle="Handbook of Mathematical Methods in Imaging",
year="2011",
publisher="Springer New York",
address="New York, NY",
pages="1016--1057",
isbn="978-0-387-92920-0"
}

@misc{NLopt,
  title = {The {NLopt} nonlinear-optimization package},
  author = {Steven G. Johnson},
  year = {2007},
  howpublished = {\url{https://github.com/stevengj/nlopt}}
}

@techreport{BOBYQA,
  author      = {M. J. D. Powell},
  title       = {The {BOBYQA} algorithm for bound constrained optimization without derivatives},
  institution = {Department of Applied Mathematics and Theoretical Physics, Cambridge University},
  year        = {2009},
  number      = {NA2009/06},
  address     = {Cambridge, UK}
}

@misc{imfusion,
  title = {{ImFusion GmbH}},
  howpublished = {\url{https://www.imfusion.com/}},
}

@INPROCEEDINGS{LNCC,
  author={Baig, Asim and Chaudhry, M. Ali and Mahmood, Azhar},
  booktitle={Proceedings of 2012 9th International Bhurban Conference on Applied Sciences \& Technology (IBCAST)}, 
  title={Local normalized cross correlation for geo-registration}, 
  year={2012},
  volume={},
  number={},
  pages={70-74}}

@ARTICLE{FFD,
  author={Rueckert, D. and Sonoda, L.I. and Hayes, C. and Hill, D.L.G. and Leach, M.O. and Hawkes, D.J.},
  journal={IEEE Transactions on Medical Imaging}, 
  title={Nonrigid registration using free-form deformations: application to breast {MR} images}, 
  year={1999},
  volume={18},
  number={8},
  pages={712-721}}

\end{document}